\documentclass[a4paper,onecolumn,11pt]{quantumarticle}
\pdfoutput=1
\usepackage[utf8]{inputenc}
\usepackage[english]{babel}
\usepackage[T1]{fontenc}
\usepackage{amsmath}
\usepackage{hyperref}

\usepackage{tikz}
\usepackage{lipsum}

\usepackage[numbers,sort&compress]{natbib}

\usepackage{amsthm}
\usepackage{graphicx}
\newcommand{\<}{\langle}
\renewcommand{\>}{\rangle}
\renewcommand{\(}{\left(}
\renewcommand{\)}{\right)}
\renewcommand{\[}{\left[}
\renewcommand{\]}{\right]}
\newcommand{\e}{\text{e}}
\newcommand{\Tr}{\text{Tr}}
\newcommand{\OO}{\mathcal{O}}
\newcommand{\CC}{\mathcal{C}}
\newcommand{\UU}{\mathcal{U}}
\newcommand{\EE}{\mathcal{E}}
\newcommand{\II}{\mathcal{I}}

\begin{document}

\title{Faster Probabilistic Error Cancellation}

\author{Yi-Hsiang Chen}
\affiliation{Quantinuum, 303 South Technology Court, Broomfield, Colorado 80021, USA}
\email{yihsiang.chen@quantinuum.com}

\maketitle

\begin{abstract}
Probabilistic error cancellation (PEC) is a leading quantum error mitigation method that provides an unbiased estimate, although it is known to have a large sampling overhead. In this work, we propose a new method to perform PEC, which results in a lower sampling cost than the standard way. It works by decomposing the inverse channel of each gate or each circuit layer into the identity part and the non-identity part and reorganizing the full circuit as different powers of the inverse generator. The ideal circuit becomes a linear combination of noisy circuits with different weights where shots are deterministically allocated to each circuit based on its weight. This naturally sets the achievable bias given a finite amount of shots. As the number of shots is increased, smaller bias terms can be gradually resolved and become bias-free in the limit of sufficient shots. We show the saving both analytically and numerically over the standard PEC and identify situations where it can outperform heuristic approach, such as zero-noise extrapolation, due to the well-controlled bias. We also demonstrated this method experimentally and found excellent agreement between the mitigated and the ideal values. 
\end{abstract}

\section{Introduction}
Quantum error mitigation has been an essential part of near-term quantum computing where it helps to recover the correct answer even in the presence of hardware errors \cite{Kim2023,floquet2025}. This generally is done by performing extra noisy circuits (with ancillary qubits in some cases) to reduce the effect of errors and inevitably comes with some overhead in the total resources required. It is a common belief that quantum error mitigation has exponential overhead \cite{caiRMP2023} as errors are not corrected and the noiseless signal should decay exponentially with the circuit depth. However, it plays a crucial role in enhancing the reliability of current small-to-intermediate scale computations and is still expected to be important even when fault-tolerant quantum error correction becomes available \cite{zim2025,aharonov2025}.

Existing error mitigation methods can be roughly categorized as either biased or unbiased ones. A biased error mitigation means the method converges to a value that is not exactly the same as the ideal noiseless value, and the distance between them is called the bias. An unbiased error mitigation means the bias is zero. Probabilistic Error Cancellation (PEC) and Zero-Noise Extrapolation (ZNE) are arguably the most representative unbiased and biased methods respectively \cite{temme2017,Li2017}. ZNE works by amplifying the error rate to learn how a noisy observable responds to noise magnitude and extrapolating to the zero noise value using some ansatz function. Since the true function underlying how the observable behaves with the noise magnitude is generally complicated and inaccessible, a bias will occur when the ansatz differs from the actual decay function. Although ZNE has been shown to work well with an exponential decay ansatz \cite{Giurgica2020,Cai2021,Kim2023,floquet2025}, it still does not provide any accuracy guarantee even assuming the ZNE protocol is carried out perfectly. On the other hand, PEC can provide an unbiased estimate. It works by implementing the inverse of the error channel to directly negate the effect of errors. However, the inverse channel is not a physically implementable operation. It can nonetheless be represented as a linear combination of implementable operations where the pseudo-ensemble of circuits recovers the ideal noiseless circuit. This unfortunately comes with an added overhead in the number of samples required to reach the same level of statistical noise as a bare noisy value \cite{temme2017}. There has been previous work on reducing the overhead by finding the optimal decomposition of the inverse channel into implementable operations \cite{Piveteau2022,Takagi2021,Guo2023,Jiang2021physical,Regula2021operational}. There are also hybrid PEC-ZNE methods that aim to combine the efficiency of ZNE and the accuracy of PEC to achieve a better performance \cite{Mari2021,McDonough2022}. For circuits that are dominated by Clifford gates, \cite{scheiber2025reducingpecoverheadpauli} shows that the PEC overhead can be reduced by propagating Pauli errors through the circuits. For local observables, one can exploit the lightcone of the observable to only include the relevant gates and reduce the overhead \cite{tran2023localityerrormitigationquantum,eddins2024lightconeshadingclassicallyaccelerated}.

In addition to the above, a central motivation of this paper is that the effect of bias is relative to the statistical noise one can achieve. Specifically for PEC, rather than aiming for completely bias-free, one should pursue a bias that is only much smaller than the achievable statistical noise. To achieve this goal, we propose a new protocol to perform PEC in a different representation. Instead of sampling an operation per gate or per layer in a circuit, we decompose each inverse channel into the identity part and the non-identity part and reorganize the circuit as a sum of different powers of the inverse generator. This allows for a systematic and deterministic way to allocate shots to different noisy circuits based on the corresponding weights, resulting to a more efficient PEC procedure with a natural control on the bias.

We begin by a brief explanation of the sub-optimality of the standard PEC protocol in Sec.~\ref{sec:optimality} and introduce the main idea of binomial expansion for PEC in Sec.~\ref{sec:main_result}. Numerical comparison of the performance of different methods are provided in Sec.~\ref{sec:numerics}. Finally, we demonstrate this method experimentally in Sec.~\ref{sec:experiment}.

\section{The sub-optimality of the overhead in PEC} \label{sec:optimality}
It was shown in \cite{kento_optimal2023} that a fundamental lower bound on the sampling cost for any unbiased error mitigation protocol is $ \propto (1+\epsilon)^l$, where  $\epsilon$ is the error rate and $l$ is the circuit depth (or gate counts). This implies the number of samples required to maintain the same statistical noise is a factor of $\propto (1+\epsilon)^{2l}$ more. It is also known that the overhead in standard PEC is sub-optimal, i.e., $\gamma_{PEC}\approx(1+2 \epsilon)^{l}$ \cite{temme2017,Suzuki2022}, implying the required shots is a factor of $(1+2\epsilon)^{2l}$ more. Here we briefly explain the origin of this sub-optimal overhead.

Let us consider an error channel $\Lambda=(1-\epsilon)\mathcal{I}+\epsilon \mathcal{E}$ attached to an ideal unitary operation $\UU$. PEC aims to perform $\Lambda^{-1}$ for every noisy operation $\Lambda \UU$ such that the sequence recovers the ideal noiseless values. Note that the inverse channel is $\Lambda^{-1}=(1+\epsilon)\II -\epsilon \EE+\OO(\epsilon^2)$, which can be checked by $\Lambda^{-1}\Lambda=\II$. Since there is a minus sign in front of the error map $\EE$, it does not correspond to a valid physical quantum channel in general (even if $\EE$ does). Therefore, PEC effectively implements this pseudo channel by performing a quasi-probability sampling that with probability $(1+\epsilon)/(1+2\epsilon)$ one implements nothing and with probability $\epsilon/(1+2\epsilon)$ one implements the error map $\EE$ while recording a minus sign when evaluating the observable. However, the observable has to be multiplied by the renormalization factor $1+2\epsilon$. Implementing this inverse channel for $l$ operations results in a factor of $(1+2\epsilon)^l$ increase in the observable, which is quadratically worse than the optimal $(1+\epsilon)^l$. It is shown in \cite{kento_optimal2023,tsubouchi2025} that the optimal scaling can be saturated with a global depolarizing error where the quantum state is gradually replaced by the global identity operator and the noisy observable simply becomes the noiseless value with an exponentially attenuated factor. However, this holds only for the global depolarizing error as the noisy value can be more complicated than a simple scaling of the noiseless value when the error channel is structured. In the following, we show that PEC's cost can be reduced using a different representation for the PEC protocol, without extra assumptions on the error channel or the circuit structure.

\section{PEC with binomial expansion}\label{sec:main_result}
Here, we describe an error mitigation strategy that effectively inverts the error channels to achieve a noiseless value. Unlike the standard PEC \cite{temme2017} which performs the inverse error channel by pseudo-probability sampling for each gate (or each circuit layer), we instead separate each inversion channel into the identity part and an error map and reorganize the sequence of operations in terms of different powers of the error maps. The noiseless observable can then be expressed as a linear combination of noisy observables where the resulting overhead is lower than the standard PEC.

Given a noisy circuit $\mathcal{C}$ consisting of $l$ noisy gates, i.e., $\mathcal{C}=\Lambda \UU_l\cdots \Lambda \UU_1$, where $\UU_i$ are the ideal gates and the error channel for each gate is $\Lambda=(1-\epsilon)\mathcal{I}+\epsilon\EE'$. We aim to perform the inverse map $\Lambda^{-1}$ for each noisy gate such that $\mathcal{C}_{ideal}=\Lambda^{-1}\Lambda\UU_l\cdots \Lambda^{-1}\Lambda\UU_1=\UU_l\cdots\UU_1$ is the target noiseless circuit. We first note that the inverse channel has a particular form
\begin{align}
\Lambda^{-1}=(1+\epsilon_1)\mathcal{I}-\epsilon_2\EE,
\end{align}
where $\epsilon_1,\epsilon_2\approx \epsilon +\OO(\epsilon^2)$ are approximately the same size of the error strength $\epsilon$ in $\Lambda$. We call $\EE$ the inverse generator. Here we assume $\EE$ is a linear combination of implementable operations, i.e., $\EE=\sum_{i} c_iV_i(\cdot)V_i^{\dagger}$ where $c_i$ are real (but not necessarily positive), $\sum_{i } |c_i|=1$ and each $V_i$ is a tensor product of single-qubit unitaries. If $\Lambda$ is a stochastic Pauli channel, then $V_i$ are non-identity Pauli operators and $\epsilon_1, \epsilon_2$ and $c_i$ can be computed straightforwardly by inverting the diagonal matrix $\Lambda$ in the Pauli-Transfer-Matrix representation \cite{greenbaum2015}. To effectively recover the noiseless circuit $\CC_{ideal}$, we expand every $\Lambda^{-1}$ as a sum of the identity map $\mathcal{I}$ and the inverse generator $\EE$ and reorganize the sequence in terms of the number of $\EE$s occurring in the sequence, i.e.,
\begin{align}
\mathcal{C}_{ideal}=\sum_{k=0}^l {l \choose k} (1+\epsilon_1)^{l-k}(-\epsilon_2)^{k}\CC_{k}, \label{eq:binom_circ}
\end{align}
where $\CC_{k}$ is the \emph{noisy} circuit involving $k$ inverse generators $\EE$ injected averaging over all possible places that the $k$ $\EE$s can occur, i.e., 
\begin{align}
\CC_{k}=\frac{\sum_{S\in\mathbf{S}_k}\CC_{S}}{{l \choose k}},
\end{align}
where $\mathbf{S}_k$ is the set of all possible sets of $k$ different locations chosen from $l$ total available positions and $S$ is a particular set of $k$ different locations. For example, $\CC_{1}=(\Lambda\UU_l\cdots\EE\Lambda\UU_1+\cdots+\EE\Lambda\UU_l\cdots\Lambda\UU_1)/l$, where $\EE$ is added at locations from the first noisy gate $\Lambda\UU_1$ to the last noisy gate $\Lambda\UU_l$. 

To obtain an unbiased estimator of an observable $\<O\>_{ideal}=\Tr\[O\CC_{ideal}(\rho)\]$, we implement each circuit $\CC_{k}$ in Eq.~\eqref{eq:binom_circ}, measure $O$ and combine them with the corresponding coefficients. Hence, the estimator is
\begin{align}
\<O\>_{est}=\sum_{k=0}^l \gamma_k \<O\>_k,  \label{eq:obs0}
\end{align}
where 
\begin{align}
\gamma_k={l \choose k}(1+\epsilon_1)^{l-k} (-\epsilon_2)^{k}\ \  \text{and}\ \ \<O\>_k=\Tr\[O\CC_{k}(\rho)\]. \nonumber
\end{align}
The value $\<O\>_k$ is an average over all possible $k$ locations to inject $\EE$ where each $\EE=\sum_{i}c_iV_i(\cdot)V_i^{\dagger}$ is implemented by applying a $V_i$ with probability $|c_i|$ while recording the sign $\text{sign}(c_i)$, i.e.,  
\begin{align}
  \<O\>_k=\frac{1}{{l \choose k}}\sum_{S\in\mathbf{S}_k}\sum_{i_k,\cdots, i_1}\text{sign}(c_{i_k})\cdots\text{sign}(c_{i_1}) |c_{i_k}|\cdots|c_{i_1}|\Tr\[O\CC_{(S,\vec{i})}(\rho)\],
\end{align}
where $\mathbf{S}_k$  is the set of all possible sets of $k$ different locations chosen from $l$ total available positions and $\CC_{(S,\vec{i})}$ is the circuit with $k$ unitaries $(V_{i_k},\dots,V_{i_1})$ injected at positions $S$. For example, if $k=2$ and $l=4$, one instance is $\CC_{(S,\vec{i})}=\mathcal{V}_{i_2}\Lambda\UU_4 \Lambda\UU_3 \mathcal{V}_{i_1}\Lambda\UU_2 \Lambda\UU_1$, where $\vec{i}=(i_1,i_2)$ is the indices of the sampled unitaries $\mathcal{V}_{i}(\cdot)=V_{i}(\cdot)V_i^{\dagger}$ and $S$ means the locations are after the second and the fourth gates. In practice, the number of locations ${l \choose k}$ can be very large that directly averaging over those circuit values $\Tr\[O\CC_{(S,\vec{i})}(\rho)\]$ requires too many circuits to be run, one can instead sample uniformly $k$ locations to insert $\EE$. Hoeffding's inequality then guarantees a fast convergence on $\<O\>_k$ with these sampled circuits since the observable $O$ is bounded. After combining every term together, one can verify that $\<O\>_{ideal}=\<O\>_{est}$ in the infinite shots limit.

\begin{figure}[h]
\centering
\includegraphics[width=8cm]{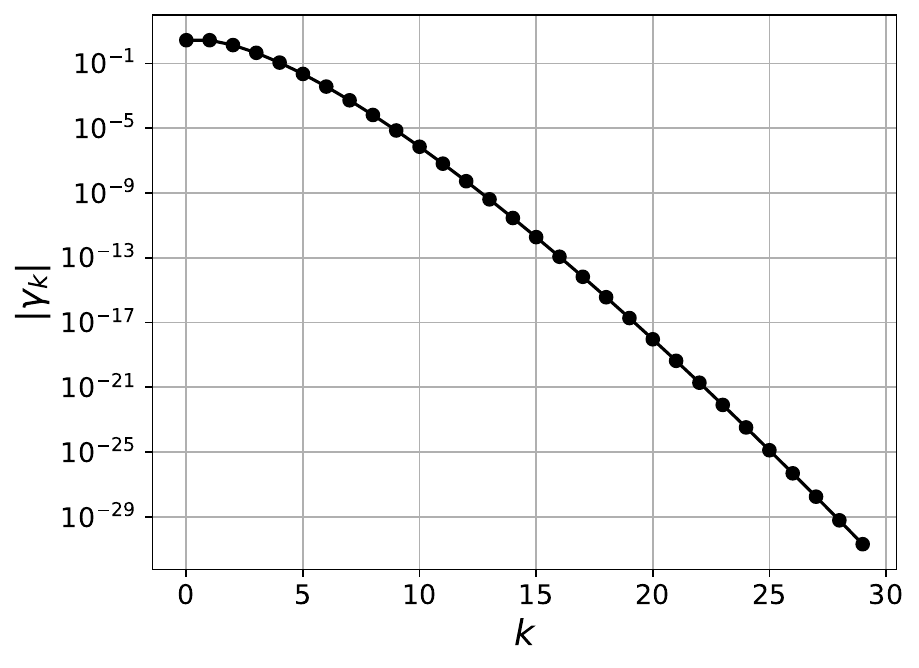}
\caption{The coefficients $|\gamma_k|$ as a function $k$, with $l=1000$ total number of gates and error rate $\epsilon=10^{-3}$ under a depolarizing error channel $\Lambda$.} 
 \label{fig:gamma_k}
\end{figure}

As shown in Fig.~\ref{fig:gamma_k}, the coefficients $\gamma_k$ decay below numerical precision very quickly under the parameter regimes where PEC's cost remains practical, i.e., when $\epsilon l=\OO(1)$. This suggests one only needs to measure the expectation values in Eq.~\eqref{eq:obs0} up to an order that is much smaller than $l$. An error mitigation technique introduced in \cite{Etienne2025} uses a linear approximation in the lower error regime, which resembles the $K=1$ truncation here. In practice, one can determine the truncation order $K$ by either the achievable statistical error with a given amount of shots or the user-defined tolerable bias. We explain the former as follows. Given $M$ shots, one allocates the shots to each $\<O\>_k$ based on the size of its coefficient $\gamma_k$, i.e., using $M|\gamma_k|/\sum_j|\gamma_j|$ shots for $\<O\>_k$. One drops the $\<O\>_k$ if the shots allocated to it is below one, i.e., when $M|\gamma_k|/\sum_{j=0}^l|\gamma_j|<1$. This means such values $\<O\>_k$ have weights $\gamma_k$ that are too small to be resolvable using $M$ shots. Collecting all the $\<O\>_k$ such that the shots allocated to each is more than one, we have our estimator truncated at order $K$, i.e., $\<O\>_{est}=\sum_{k=0}^K \gamma_k \<O\>_k$ where we allocate $M|\gamma_k|/\sum_{j=0}^K|\gamma_j|$ shots to $\<O\>_k$. In addition, one can also manually truncate the series to order $K$ such that the residual terms do not contribute more than a tolerable bias $\delta$. Specifically, given a bias tolerance $\delta$, one can find a $K$ such that $||O|| \sum_{k=K+1}^l |\gamma_k| \leq \delta$, where $||\cdot||$ is the operator norm. Indeed, the bias from truncating at order $K$ can be bounded by $|\sum_{k=K+1}^l \gamma_k \<O\>_k|\leq \max_{k}|\<O\>_k| \sum_{k=K+1}^l |\gamma_k|\leq||O|| \sum_{k=K+1}^l |\gamma_k| \leq \delta$. For a normalized observable $||O||=1$, the bias at truncation $K$ is bounded by $\sum_{k=K+1}^l |\gamma_k|$ which can be numerically computed efficiently.
To summarize the protocol,
\begin{enumerate}
\item given the error channel $\Lambda$, find its inverse $\Lambda^{-1}=(1+\epsilon_1)\mathcal{I}-\epsilon_2\EE$
\item given $M$ shots, find the truncation order $K$ by either the shot-limited truncation or the user-defined bias tolerance $\delta$
\item allocate $M|\gamma_k|/\sum_{j=0}^K|\gamma_j|$ shots to each observable $\<O\>_k$
\item measure each $\<O\>_k$ by sampling $k$ locations uniformly at random from total $l$ positions and for each location applying a unitary $V_{i}$ with probability $|c_i|$ and recording the sign $\text{sign}(c_i)$
\item output the estimator as $\<O\>_{est}=\sum_{k=0}^K \gamma_k \<O\>_k$
\end{enumerate}

\subsection{Resource Estimation and Comparison}
Here we evaluate the cost of this mitigation protocol. Recall that the estimator $\<O\>_{est}$ is a linear combination of different noisy observables $\<O\>_k$ where each observable is allocated with $M|\gamma_k|/\sum_{j=0}^K|\gamma_j|$ shots. The variance of the estimator using a total $M$ shots is given by the sum of $\gamma_k^2$ multiplying the variance of each $\<O\>_k$ using $M|\gamma_k|/\sum_{j=0}^K|\gamma_j|$ shots, i.e.,
\begin{align}
&\frac{\text{Var}[\<O\>_{est}]}{M}=\sum_{k=0}^K \frac{\gamma_k^2 \text{Var}\[\<O\>_k\] }{M|\gamma_k|/\sum_{j=0}^K|\gamma_j|},
\end{align}
where $\text{Var}[\<O\>_{est}]$ represents the variance per shot for the estimator. Therefore the estimator variance is 
\begin{align}
\text{Var}[\<O\>_{est}]=\(\sum_{k=0}^K|\gamma_k|\)\sum_{k=0}^K|\gamma_k|\text{Var}\[\<O\>_k\]. \label{eq:est_var}
\end{align}
Now we explain why this variance is lower than that of the standard PEC. The saving is two-fold---the smaller overhead factor due to the truncation and the deterministic allocation of shots to each circuit.
To see the saving from the series truncation $K$, assuming the variance of each $\<O\>_k$ is similar for all $k$, the variance becomes $\text{Var}[\<O\>_{est}]=\(\sum_{k=0}^{K}|\gamma_k|\)^2\text{Var}[\<O\>]\leq(\sum_{k=0}^l |\gamma_k|)^2\text{Var}[\<O\>]=(1+\epsilon_1+\epsilon_2)^{2l}\text{Var}[\<O\>]\approx (1+2\epsilon)^{2l}\text{Var}[\<O\>]$, where the last expression is the variance of the standard PEC. This saving depends on the truncation order $K$. Suppose we truncate at the zeroth order $K=0$, then we have $\<O\>_{est}=(1+\epsilon_1)^l\<O\>_0$ with the variance overhead $(1+\epsilon_1)^{2l}$ which saturates the lower bound \cite{kento_optimal2023}. Such lower bound can be achieved when the error channel $\Lambda$ is a global depolarizing channel. Indeed, a global depolarizing error replaces the state as the identity state and the identity state does not change with any unitary operator. Hence, any circuit with one or more error map injected replaces the state as the identity state and we have $\<O\>_k=0$ for all $k\geq1$ when $O$ is any Pauli observable. 

The second reason for the lower cost of $\<O\>_{est}$ comes from the deterministic allocation of shots to each observable $\<O\>_k$ based on the weights $|\gamma_k|/\sum_{j}|\gamma_j|$ as opposed to \emph{sampling} each $\<O\>_k$ with probability $|\gamma_k|/\sum_{j}|\gamma_j|$, e.g., as described in \cite{caiRMP2023}, which comes with an extra variance from the difference between the values $\<O\>_k$. To show this, we first define $\gamma:=\sum_{k}|\gamma_k|$ and the probability distribution $p_k:=|\gamma_k|/\gamma$. The variance of the estimator is $\text{Var}[\<O\>_{est}]=\gamma^2\sum_k p_k\text{Var}[\<O\>_k]$ in Eq.~\eqref{eq:est_var}. On the other hand, if one constructs the estimator $\<O\>_{sampling}:=\gamma\sum_k p_k\text{sign}(\gamma_k)\<O\>_k$ by sampling and measuring $\<O\>_k$ with probability $p_k$ for each shot \footnote{In the standard PEC procedure \cite{temme2017}, each $\<O\>_k$ is further expanded down as a quasi-probabilistic ensemble of expectation values. But for the simplicity purpose of explaining the saving, we keep it at the $\<O\>_k$ level}, then the variance is
\begin{align}
\text{Var}[\<O\>_{sampling}]&=\gamma^2\[\sum_kp_k\sum_b p(b|k) \<b|O|b\>_k^2-\(\sum_k p_k\text{sign}(\gamma_k)\<O\>_k\)^2\] \nonumber\\
&=\gamma^2\[\sum_kp_k (\text{Var}[\<O\>_k]+\<O\>_k^2)-\(\sum_k p_k\text{sign}(\gamma_k)\<O\>_k\)^2\] \nonumber\\
&=\text{Var}[\<O\>_{est}]+\gamma^2\[\sum_k p_k\<O\>_k^2-\(\sum_kp_k\text{sign}(\gamma_k)\<O\>_k\)^2\], \nonumber\\
&:=\text{Var}[\<O\>_{est}]+\Delta,  \label{eq:var_compare0}
\end{align}
where $p(b|k)$ is the probability of obtaining the bitstring $b$ on the $k$th circuit and $\<b|O|b\>_k$ is the expectation value of the measured bitstring $b$. The difference $\Delta=\text{Var}[\<O\>_{sampling}]-\text{Var}[\<O\>_{est}]$ in Eq.~\eqref{eq:var_compare0} is non-negative, i.e., $\Delta\geq 0$ implied from using Cauchy-Schwarz inequality between two vectors $u_k=\sqrt{p_k}$ and $v_k=\sqrt{p_k}\text{sign}(\gamma_k)\<O\>_k$, where $\Delta=0$ happens only when $\text{sign}(\gamma_k)\<O\>_k$ are the same for all $k$. This implies the cost of deterministically allocating the shots to each observable $\<O\>_k$ is lower than that of sampling each $\<O\>_k$ with probability $p_k$.

\section{Performance comparison}\label{sec:numerics}
Here, we compare the performance of our protocol, which we called Faster PEC (FPEC) with the standard PEC and ZNE. We consider the dynamics simulation of the two-dimensional transverse-field Ising model (2D TFIM) with periodic boundary on both dimensions, i.e.,
\begin{align}
H=\sum_{<i,j>}JZ_iZ_j+\sum_{j=1}^N hX_j, \label{Ham}
\end{align}
where $<i,j>$ indicates the nearest-neighbor pairs on a torus. The circuit is the second-order Trotterized unitary, i.e.,
\begin{align}
U_{trot}:=\prod_{j=1}^N \e^{-ih X_j \tau/2}\prod_{<i,j>}\e^{-iJ Z_iZ_j \tau}\prod_{j=1}^N \e^{-ih X_j \tau/2}, \label{U_trot}
\end{align}
where $\tau$ is the step size and $N$ is the number of qubits. We evolve the state using $r$ repetitions of $U_{trot}$ and measure an observable. The error model is a structured stochastic Pauli error channel attached to each two-qubit gate $\e^{-iJ Z_iZ_j \tau}$ and we assume there is no other error in the circuits.

\begin{figure}[h]
\centering
\includegraphics[width=8cm]{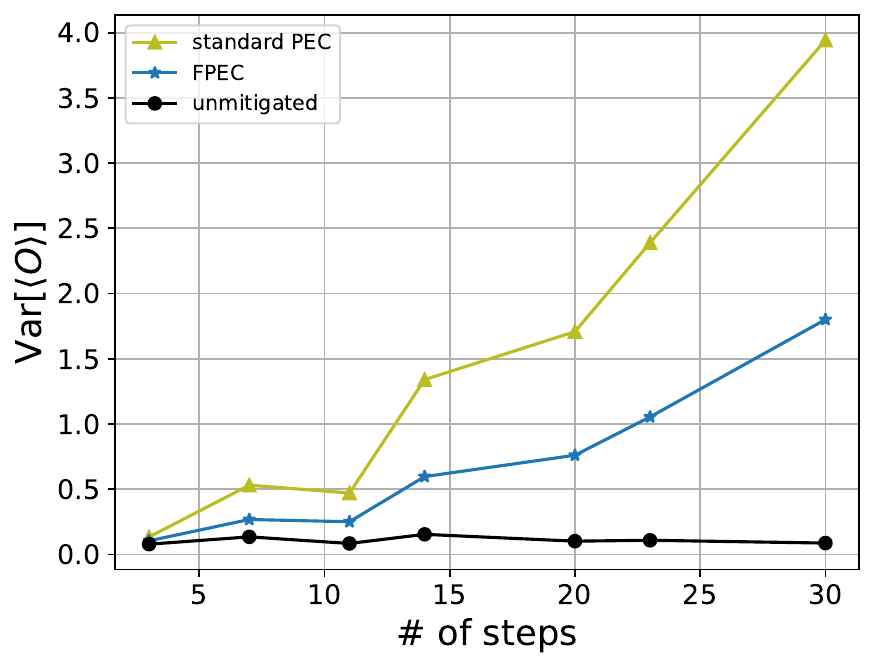}
\caption{Comparison of the variance of the raw (unmitigated) value and the mitigated values from FPEC and the standard PEC, as a function of the number of Trotter steps.}
\label{fig:pec_var}
\end{figure}

We first compare the variance between the standard PEC and FPEC. In standard PEC, we sample a Pauli channel (including the identity) from the inverse channel $\Lambda^{-1}$ with the corresponding probability and apply it after each two-qubit gate \cite{temme2017}. The signs of the sampled Paulis are recorded and combined as a single sign for the sampled circuit. This is done for every shot and all the measured values are combined with the corresponding signs and factors to obtain the estimator $\<O\>_{PEC}$. We ran total $M=5000$ shots and computed the standard deviation $\sigma_{PEC}$ of the 5000 values and deduce the variance of the PEC estimator as $\text{Var}[\<O\>_{PEC}]=\sigma^2_{PEC}$. In FPEC, we are also given the same total $M=5000$ shots and the truncation $K$ is determined by the shot-limited truncation, i.e., dropping all terms such that $M|\gamma_k|/\sum_{k=0}^l|\gamma_k|<1$. We then allocate each $\<O\>_k$ with $M|\gamma_k|/\sum_{j=0}^K|\gamma_j|$ shots for $k=0,\dots,K$ and combine them to obtain the estimator as $\<O\>_{est}=\sum_{k=0}^K\gamma_k\<O\>_k$. We randomly sample $k$ locations and apply $k$ unitaries $V_{i_k},\dots,V_{i_1}$ each sampled from the distribution $|c_i|$, for each shot allocated to $\<O\>_k$. The standard deviation of $\<O\>_k$ with $M|\gamma_k|/\sum_{j=0}^K|\gamma_j|$ shots is obtained by the standard deviation of the measured expectation values dividing by the given $M|\gamma_k|/\sum_{j=0}^K|\gamma_j|$ shots. We then use the variance of the sum relation to obtain the total variance and multiply it by $M$ to obtain the variance of the estimator $\text{Var}[\<O\>_{est}]$. The observable we measure here is $\<O\>=\<(\sum_{j=1}^NZ_j/N)^2\>$. The parameters in the simulation are $N=20$ (4-by-5 lattice), $J=1$, $h=2$, $\tau=0.2$, and two-qubit gate $\e^{-iZZ0.2}$ averaged infidelity $5.3\times 10^{-4}$. Fig. \ref{fig:pec_var} shows  $\text{Var}[\<O\>_{est}]$ can be $\sim 2.4$ times smaller than $\text{Var}[\<O\>_{PEC}]$, which means FPEC requires a factor of 2.4 less shots to obtain the same statistical error in the mitigated value.

\begin{figure}[h]
\centering
\includegraphics[width=8cm]{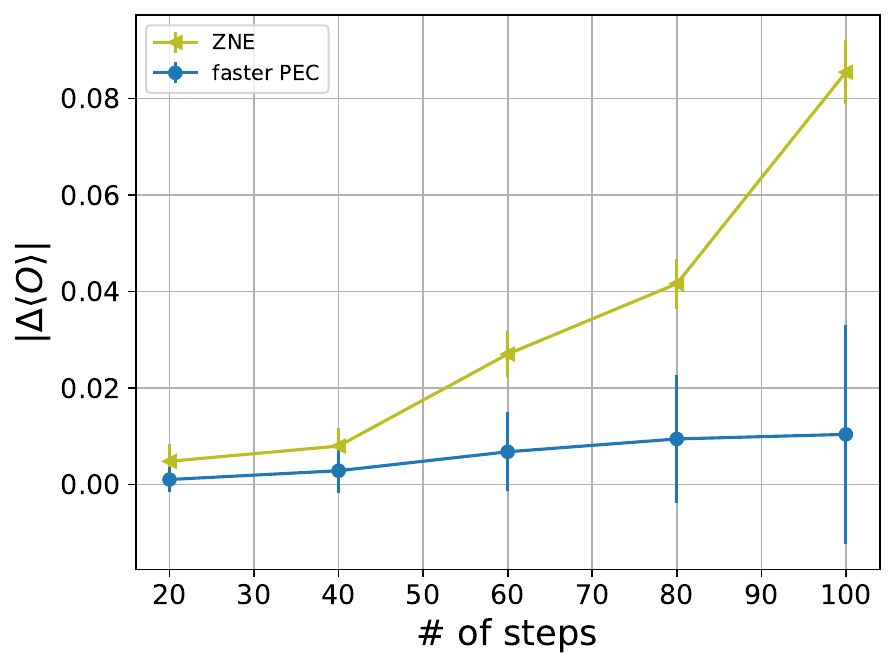}
\caption{The comparison of the absolute observable error in the mitigated value between FPEC and ZNE, as a function of the number of Trotter steps.}
\label{fig:zne_pec_bias}
\end{figure}

Now we compare the performance between ZNE and FPEC. ZNE has been a widely used error mitigation heuristic that aims to extrapolate the correct value by learning how the noisy observable scales with the error rate \cite{caiRMP2023}. Here we report cases where FPEC outperforms ZNE under the same amount of resources. Again, we simulate the 2D TFIM dynamics above and evaluate the observable $\<O\>=\<(\sum_{j=1}^NZ_j/N)^2\>$. ZNE is performed by running circuits at two-qubit error rates $6\times 10^{-4}$ and $2.4\times 10^{-3}$ and extrapolate to the zero error value using an exponential decaying ansatz. In FPEC, we output the mitigated value as $\<O\>_{est}=\sum_{k=0}^K\gamma_k\<O\>_k$ where we measure each $\<O\>_k$ using $M|\gamma_k|/\sum_{k=0}^K|\gamma_k|$ shots under a bias tolerance of $0.001$. The simulation is done in a 3-by-3 lattice with the initial state $(\cos(\pi/12)|0\>+\sin(\pi/12)|1\>)^{\otimes N}$. Fig. \ref{fig:zne_pec_bias} shows the bias, defined as the absolute difference between the mitigated value and the true value, as a function of the number of Trotter steps. Both methods are given the same amount of shots per mitigated value. FPEC is statistically consistent with no bias (i.e., the bias is due to shot noise) while the bias in ZNE increases with the circuit depth and becomes the dominant source of error. Overall, the FPEC outperforms ZNE due to the better controlled bias albeit with larger error bars. 

\section{Experimental Implementation}\label{sec:experiment}
\begin{figure}[h]
\centering
\includegraphics[width=7cm]{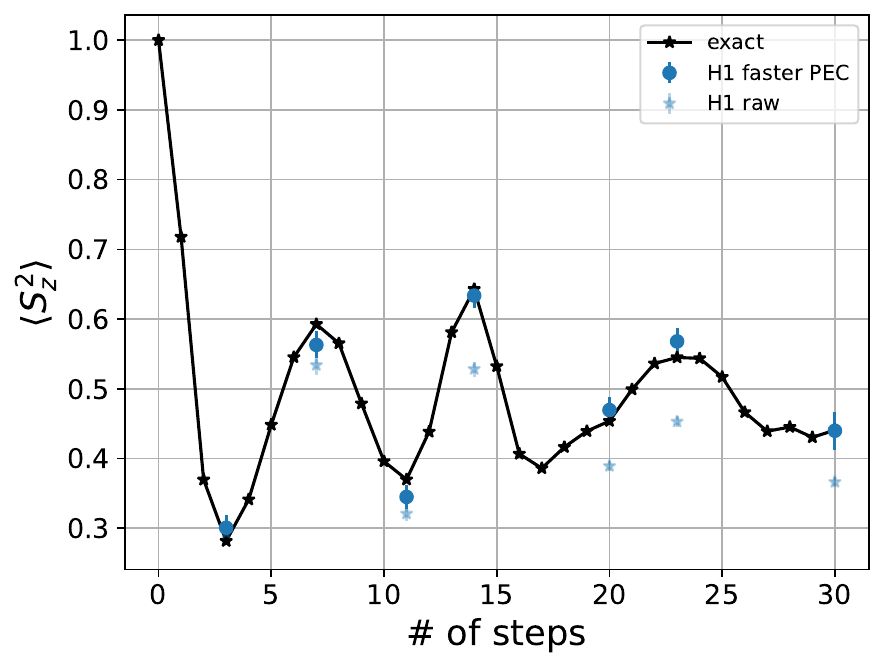}
\includegraphics[width=7cm]{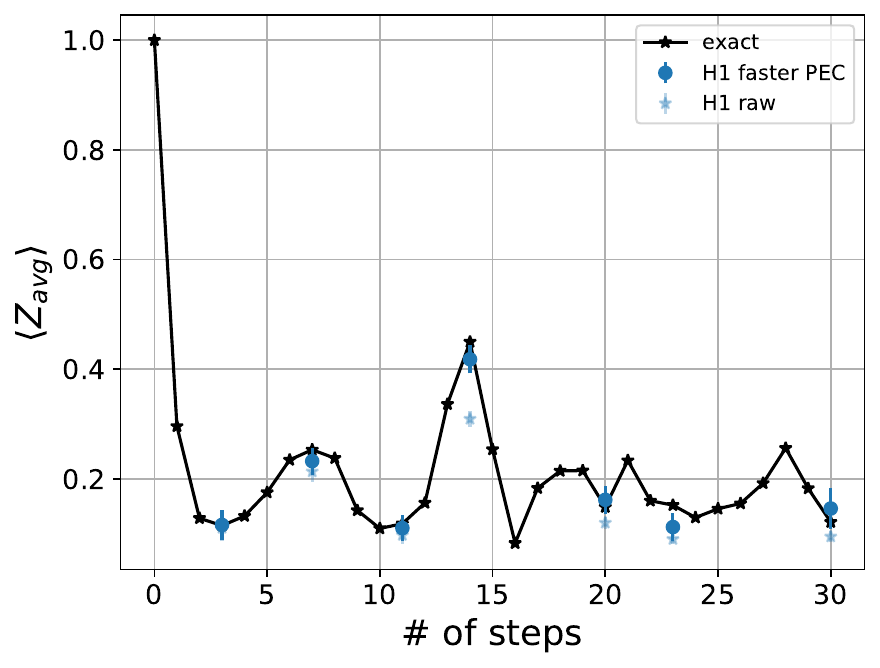}
\caption{Left: The expectation value $\<S_z^2\>:=\<(\sum_{j=1}^NZ_j/N)^2\>$ as a function of the number of Trotter steps. Right: The expectation value $\<Z_{avg}\>:=\<\(\sum_{j=1}^N\prod_{i=1}^j Z_i\)/N\>$ as a function of the number of Trotter steps. ``H1 faster PEC'' represents FPEC protocol and ``H1 raw'' represents the bare noisy values. ``Exact'' represents the correct noiseless values.} 
 \label{fig:h1_values}
\end{figure}

Here, we demonstrate FPEC protocol on Quantinuum's H1 quantum processor with twenty qubits. We simulate the 2D TFIM dynamics with Eqs.~\eqref{Ham} and \eqref{U_trot} on a 4-by-5 lattice. The parameters are $J=1$, $h=2$, $\tau=0.2$ and the initial state $|0\rangle^{\otimes N}$. We first measure the infidelity of the non-Clifford two-qubit gate $\e^{-iZZ0.2}$ using a direct randomized benchmarking method \cite{proctor_directRB2019,H2paper2023} and compute the inverse channel $\Lambda^{-1}$ assuming the two-qubit gate's error channel $\Lambda$ is a depolarizing error \footnote{This assumption may not hold perfectly for the two-qubit gate we use. One technical difficulty for characterizing the error channel is the gate is non-Clifford, where the standard twirling used in Cycle-Benchmarking for Clifford gates breaks down.  More detailed error model characterization is still possible, but not without further assumptions (e.g., assuming $\e^{\pm iZZ0.2}$ have exactly the same error channel) or overhead \cite{layden2025theoryquantumerrormitigation}. Furthermore, using a depolarizing error channel as an approximation is a mean to test the robustness of this error mitigation method to the error model mischaracterization. The experimental results indicate this method is fairly robust to this inperfection}. In addition, we implement $X$ pulses for each two-qubit gate in each Trotter layer to minimize the effect of coherent errors in the form of Z-type rotations. The PEC protocol is carried out by setting a bias tolerance of $0.01$ which provides the truncation order $K$ for the circuit. In the circuits we ran, $K$ increases from 1 to 4 with the circuit depths. We measure each observable $\<O\>_k$ by uniformly sampling $k$ two-qubit gate positions to add Pauli gates for the $k$ chosen two-qubit gates and combine the noisy observables to obtain the mitigated value $\<O\>_{est}=\sum_{k=0}^K\gamma_k\<O\>_k$. Fig.~\ref{fig:h1_values} shows excellent agreement between the mitigated values and exact noiseless ones. We note that although the underlying two-qubit gate error model may not be exactly depolarizing and the inversion may not cancel the error completely, the protocol still shows a fairly robust tolerance to the imperfect channel characterization.

\section{Conclusion}

We proposed a new error mitigation protocol that virtually implements the inverse of the error channel such that the noiseless value is recovered as a linear combination of noisy values. This method exploits the fact that the inverse of the error channel is close to the identity operation in the lower error regime. By re-grouping terms in different powers of the inverse generator, the noiseless circuit can be expressed as a linear combination of noisy circuits where most of the circuits have weights that are negligibly small. This representation allows for an intuitive control on the bias to be below statistical relevance. This protocol is more efficient than the standard PEC in terms of the total sampling cost. The saving comes from a) the truncation of the series and b) the deterministic shots allocation to each noisy observable. We test the saving numerically on a 2D TFIM circuits and observe up to $\sim 50\%$ less shots required to achieve the same statistical error. We note that the savings can vary from circuits to circuits. The saving from a) depends on the aggressiveness of the truncation, e.g., a prior knowledge of the observable decay can help reduce the number of noisy circuits to be implemented. The saving from b) is higher the more the noisy observables $\<O\>_k$ differ from each other. We also explore cases where a biased error mitigation method like ZNE performs worse than this new PEC protocol under the same amount of resources. This typically happens when the observable error is dominated by the ZNE bias rather than the statistical noise. Of course there are numerous other error mitigation methods in the literature (e.g., \cite{caiRMP2023}) that each may be advantageous in certain cases. Instead of comparing the performance against every one of them, it may be more beneficial to explore hybrid methods that combine the advantage of each since different saving techniques may coexist and the protocol provided in this paper can improve the cost of the PEC part of a hybrid method.

\begin{acknowledgments}
We thank Etienne Granet, Christopher Self, Karl Mayer and Charles Baldwin for valuable feedback and Quantinuum's H1 team for carrying out the experiment.
\end{acknowledgments}

\bibliographystyle{plainnat}
\bibliography{rf.bib}




\end{document}